# General Invariant Velocity Originated from Principle of Special Relativity and Triple Special Theories of Relativity

Peng-Cheng Zou[1]    Yong-Chang Huang[2,3]
1. Department of Physics, Sichuan University, Chengdu, 610064, China
2. Institute of Theoretical Physics, Beijing University of Technology, Beijing, 100124, China
3. CCAST ( World Lab. ), P. O. Box 8730, Beijing, 100080, China

Abstract

This Letter, for the first time, proves that a general invariant velocity is originated from principle of special relativity, namely, discovers origin of the general invariant velocity. When the general invariant velocity is taken as light velocity in current theories, we get the corresponding special theory of relativity. Further, this Letter deduces triple special theories of relativity in cosmology, and cancels the invariant presumption of light velocity. When a kind of matter with the maximally invariant velocity that may be superluminal or equal to light velocity is determined by experiments, then utilizing this Letter's theory, all results of current physical theories are consistent.

1. Introduction

Now the cosmology is genuinely studied by scientific methods [1, 2], it is well known that, for the ultra-early universe, some theories about `extreme physics' may yield many universes that sprout from the separative big bangs into the disjoint regions of spacetime [3, 4]. Ref.[4] pointed out: if superstrings (or some other equally comprehensive theory) were `battle tested' by the convincingly explained things we could observe, then if it predicts multiple universes we should take them seriously too, just as we give credence to what our current theories predict about quarks inside atoms.

Linde[5] described a version of chaotic inflation which leads to a division of the universe into infinitely many open universes with all possible values of Omega from 1 to 0. In general, different universes in the cosmology (or multiverse) may have different physical laws, and it is well known that different universes may have different physical constants and different primordial and boundary conditions. For the different physical laws, how can we do the research ? Usually, researchers of physics start to investigate new physics in very new fields from the known special physical laws and facts to the general situations.

Relativity is one of the important bases of modern physics, e.g., see Ref.[6]. In general relativity，i．e., in 4-dimensional pseudo－Riemann manifold, one can have local flat space by taking local normalized frames of the manifold[7], in the local spacetime all results of special theory of relativity keep to stand[8]. Therefore, regardless that there is or isn't gravity, special theory of relativity is always very important. The necessity of the linearity of relativistic transformations between inertial systems was proposed[9], and the spacetime exchange invariance is given[10]. Up to now there still exists a basic problem criticized about the invariant principle of light velocity．A．Einstein in his book[11] admitted: The theory of relativity is often criticized for giving, without justification, a central theoretical role to the propagation of light．In order to give physical significance to the concept of time，processes of some kinds are required which enable relations to be established between different places．It is immaterial what kind of processes one chooses for such a definition of time．And he said：It is advantageous, however, for the theory, to choose only those processes concerning which we know something certain．This holds for the propagation of light in vacuums in higher degree than for any other process which could be considered．

Einstein's words above can not stop the criticism about the problem of invariant property of light velocity．And even some ones think that the now experiments have not rather exactly proved the invariance of light velocity. Therefore building a general theory of special relativity without the invariant assumption of light velocity is essential for solving the problem about the invariance of light velocity．Ref.[12] proposes a new kinematical derivation of the Lorentz transformations and the particle description of light, Relativity without light is given[13], Ref.[14] gave Reciprocity principle and Lorentz transformations, Wheeler and Feynman studied classical electrodynamics in terms of direct particle action satisfying causal relation[15].

This Letter is to build a general theory of special relativity of the cosmology (or multiverse), only by means of the principle of special relativity in general situation. Because of no considering the spacetime effect of gravity field in special relativity, then, the space is homogeneity, therefore, using the space homogeneity is not to belong to introduce a new presumption as in special theory of relativity[11], and this Letter needn't introduce the invariance of light velocity. In this Letter, it is proved that there exists a general constant K, for K > 0, = 0 and < 0, they give three kinds of possible relativism theories corresponding to different universes in the cosmology (or multiverse).

Before starting concrete discussions，we postulate that in any inertial coordinate system ( in the cosmology (or multiverse) ) whose coordinates and clocks at points have be defined and maintained to be synchronistical respectively, which are able to be done and don't depend on the invariant assumption of light velocity．

No losing generality, we consider the case of one dimensional space and time in the cosmology．Assume two inertial systems O and O' and their relative velocity v, consider two events a and b and their intervals （Δx，Δt） and （Δx', Δt') in O and O', respectively．

We first analyze the case of Δx' = 0, i.e., in O' watching events a and b take place at the same point，and in O watching a and b they are not at the same point ( see Fig.1），

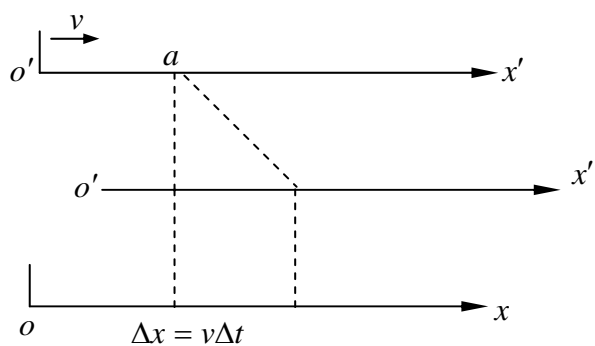

Fig.1 Relative Property of Different Points

then we have

$$\Delta x = v \Delta t \qquad (\Delta x'=0) \tag{1}$$

The above expression shows the relative property of the same point in different reference systems in the cosmology, that is, the events a and b taking place at the same point in reference system O' are watched in reference System O not to be at the same point. In fact, Eq.(1) is nothing but the definition of relative motion of two inertial reference systems in the cosmology.

Then consider the case of $\Delta t' = 0$, i.e., assume that the events a and b take place at the same time and different points in O'. Then for O, if are they at the same time? Exactly speaking, the answer should be given by experiments. In general, there exist two kinds of cases, at the same or no. For the former, it has been described by Galilean relative principle; for the latter, in terms of space homogeneity we can prove that the time difference $\Delta t$ is proportional to $\Delta x$.

Assume four events a, b, c and d, in O' they are watched at the same time and in O their coordinates x, $x + \Delta x$, $x_1$ and $x_1 + \Delta x$ respectively (see Fig. 2).

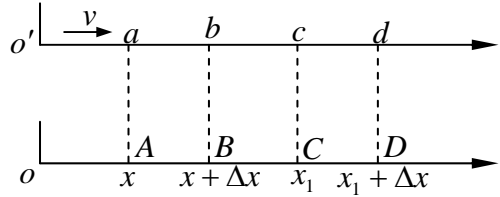

Fig.2 Relative Distribution of Different Points

If the time difference of a and b observed in O is $\Delta t$, because of the homogeneity of space, natural law has no relation with the appearing places, therefore it can be known that the time difference of d and c observed in O should also be $\Delta t$. Now taking b and c to be the same event such that distance of D and A is $2\Delta x$, thus in terms of the above discussion the time difference of D and A is $2\Delta t$. In the same deducing logic it can be proved that for two events of distance $n\Delta x$ in O, their time difference is $n\Delta t$, which just proves that for the events appearing at the same time in O' their time difference observed in O is proportional to their distance in O, that is,

$$\Delta t = f \Delta x \qquad (\Delta t'=0) \tag{2}$$

where f is the proportional constant that has nothing to do with coordinates of spacetime. Eq.(2) represents the relative property of the same time in different reference systems. In Eq.(2) when $f = 0$, which just corresponds to the case of Galilean relative principle. Since we don't give any limit on f, then Eq. (2) is general it contains all possible cases, thus it isn't a new assumption.

Starting from Eq.(1) and Eq.(2) and using the principle of special relativity, we can prove that f is proportional to the relative velocity v of two reference systems in the cosmology, the proportional constant $K = f / v$ is irrelative to the reference systems and is a general constant, and $K = 0$ is a bifurcation point, through which we can obtain three branches, which give three kinds of possible relativism theories corresponding to different universes in the cosmology (or multiverse) in which for $K > 0$ there exists a limit velocity or invariant velocity and it corresponds to the case of the special relativity in our universe.

In following section, we start from (1) and (2) and study relative properties of spacetime intervals in different reference systems in the cosmology.

## 2. Relative properties of spacetime intervals

We first discuss relative property of space distance. Assume points A, B and B" in O and A' and B' in O' (see Fig.3).

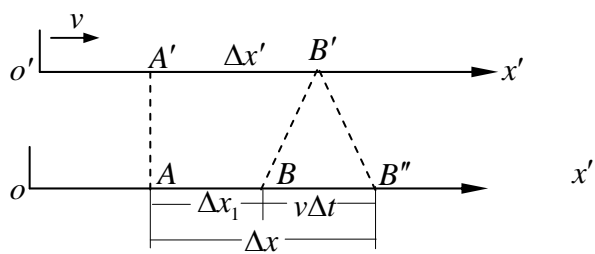

Fig．3 Relative Property of Space Distance

For O when A and A' coincide, B and B' overlap; for O' when A and A' coincide, B' coincide with B". Owing to relative property of the same time, for system O the coincidence of B' and B", the coincidence of A and A' are not at the same time, we have their time difference

$$\Delta t = f \Delta x \tag{3}$$

just in the time interval B' moves from B to B", thus

$$\Delta x_1 = \Delta x - v \Delta t = (1 - fv) \Delta x \tag{4}$$

On the other hand, the observer in O measures $\Delta x'$ to be $\Delta x_1$, conversely, $\Delta x$ is measured in O' to be $\Delta x'$. According to principle of special relativity, different inertial systems are equivalent each other. If the rule structures of measuring them in O and O' are the same，then we have

$$\frac{\Delta x_1}{\Delta x'} = \frac{\Delta x'}{\Delta x} \tag{5}$$

Solving Eqs.(4) and (5), we obtain

$$\frac{\Delta x_1}{\Delta x'} = \frac{\Delta x'}{\Delta x} = \sqrt{1 - fv} \tag{6}$$

where the minus in （6） has be taken away because it stands for the converse direction of the both coordinate systems O and O'. Eq.(6) is just Lorentz contracting formula. From the above discussions, we can see that it is the direct result from Eq. (1) and Eq. (2).

Now we study time intervals．Assume points A and B in O and A' and B' in O'（see Fig．4）.

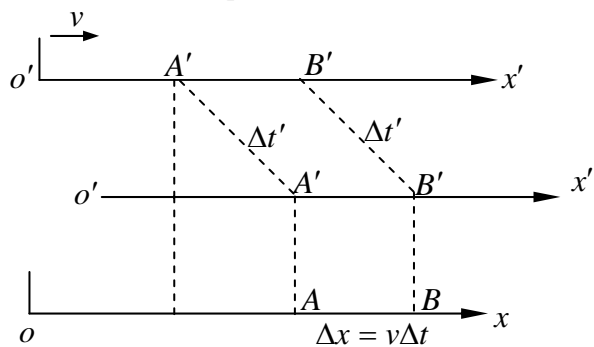

Fig.4 Relative Property of Time Intervals

At the start，B' and A coincide，for O'，after $\Delta t'$, B' and B coincide, so do A' and A．For O, from the start of coincidence of B' and A, undergoing time $\Delta t_1$, A' coincides with A, and marking the time from the coincidence of B' and A to the coincidence of B' and B is $\Delta t$. Owing to the relative property of the same time, for system O the coincidence of A and A' and the coincidence of B and B' are not at the same time. Using Eq.(2), their time difference is

$$\Delta t - \Delta t_1 = f \Delta x \tag{7}$$

Where $\Delta x$ is motion distance of B' in $\Delta t$, then we have

$$\Delta x = v \Delta t \tag{8}$$

Making use of Eqs.(7) and (8) one may obtain

$$\Delta t_1 = (1 - fv) \Delta t \tag{9}$$

On the other hand, the time of B' moving from A to B is measured in O' to be $\Delta t'$ by the clock fixed B', and in O the time is measured to be $\Delta t$ by both synchronistical clocks fixed at A and B. Conversely, the time from the coincidence of A and B' to the coincidence of A and A' is recorded in O to be $\Delta t_1$ by the clock fixed at Point A, in O' the time is synchronistically measured to be $\Delta t'$ by both clocks fixed at B' and A', According to the equivalent principle of the two inertial systems, if the clocks' structures in O and O' are identical, then we have

$$\frac{\Delta t'}{\Delta t} = \frac{\Delta t_1}{\Delta t'} \tag{10}$$

With the use of Eqs.(9) and (10), we obtain

$$\frac{\Delta t'}{\Delta t} = \frac{\Delta t_1}{\Delta t'} = \sqrt{1 - fv} \tag{11}$$

In the same reason, we neglect the minus in (11), the above expression is just the slowing formula of moving clock. Analogous to the contraction formula (6) of space length, Eq.(11) is a direct result by means of (1) and (2).

## 3. Lorentz transformation

In general, assume any two events a and b. An observer in O observes their positions at A and B their spacetime intervals ($\Delta x$, $\Delta t$); another observer in O' finds their positions at A' and B' and their spacetime intervals ($\Delta x'$, $\Delta t'$) (see Fig.5).

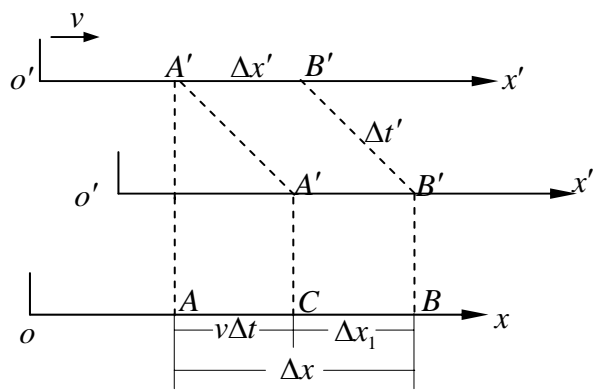

Fig.5 Relative Property of Space and Time Intervals

We first calculate the relation among $\Delta x'$, $\Delta x$ and $\Delta t$. Consider an observer in O, he sees when event a takes place, A' coincides with A, up to the appearance of event b, A' has moved from A to C, AC = $v\Delta t$. Thus the observer in O measures $\Delta x'$ to be

$$\Delta x_1 = \Delta x - v\Delta t \tag{12}$$

Using Eq.(12) and Lorentz contracting formula (6), we have the relation

$$\Delta x' = \frac{\Delta x_1}{\sqrt{1-fv}} = \frac{\Delta x - v\Delta t}{\sqrt{1-fv}} \tag{13}$$

Then we calculate the relation among $\Delta t'$, $\Delta x$ and $\Delta t$. In fig.5 we introduce event D'(see Fig.6) such that for an observer in O, he sees D' happen at the same point of B, and for another observer in O', he watches events D' and A' happen at the same time.

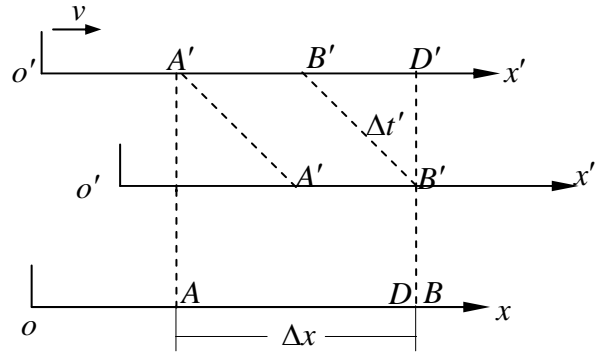

Fig.6 Relative Relation of Space and time Intervals

The time difference of events D' and B' observed in O' is $\Delta t'$ because D' and A' are at the same time watched in O'. We now find its value that is watched by an observer in O. Because of relative property of the same time, an observer in O sees D and A not be at the same time, using Eq.(2) the time difference of D and A is $f\Delta t$. Therefore the time difference of D and B is

$$\Delta t_1 = \Delta t - f\Delta x, \tag{14}$$

which is measured by a clock fixed at B point in O. With the use of Eqs.(11) and (14), we obtain

$$\Delta t' = \frac{\Delta t_1}{\sqrt{1-fv}} = \frac{\Delta t - f\Delta x}{\sqrt{1-fv}} \tag{15}$$

Eqs.(13) and (15) are nothing but Lorentz transformation formulas, and we see that the relative property originating from principle of special relativity is also a representation of spacetime uniformity.

In following section, we find the relation between f and v and formulas of velocity addition.

## 4. Formulas of velocity addition and a general constant

Assume three inertial system O, O' and O" in an universe in the cosmology (or multiverse), the velocities of O' and O" respectively relative to O and O' are $v_1$ and $v_2$, and the spacetime intervals of two events a and b observed in the three inertial systems are ($\Delta x$, $\Delta t$), ($\Delta x'$, $\Delta t'$) and ($\Delta x"$, $\Delta t"$) respectively.

Using Eqs.(13) and (15), we have

$$\begin{cases} \Delta x' = \dfrac{\Delta x - v_1 \Delta t}{\sqrt{1 - f_1 v_1}} \\ \Delta t' = \dfrac{\Delta t - v_1 \Delta x}{\sqrt{1 - f_1 v_1}} \end{cases} \quad (16)$$

and

$$\begin{cases} \Delta x'' = \dfrac{\Delta x' - v_1 \Delta t'}{\sqrt{1 - f_1 v_1}} \\ \Delta t'' = \dfrac{\Delta t' - v_1 \Delta x'}{\sqrt{1 - f_1 v_1}} \end{cases} \quad (17)$$

Substituting Eqs.(16) into Eq.(17), Eq. (17), thus, can be rewritten as

$$\begin{cases} \Delta x'' = \dfrac{1 + f_1 v_2}{\sqrt{1 - f_1 v_1}\sqrt{1 - f_2 v_2}} (\Delta x - \dfrac{v_1 + v_2}{1 + f_1 v_2}) \Delta t \\ \Delta t'' = \dfrac{1 + f_2 v_1}{\sqrt{1 - f_1 v_1}\sqrt{1 - f_2 v_2}} (\Delta t - \dfrac{f_1 + f_2}{1 + f_2 v_1}) \Delta x \end{cases} \quad (18)$$

On the other hand, assume $v_3$ to be the velocity of O" relative to O, then we have

$$\begin{cases} \Delta x'' = \dfrac{1}{\sqrt{1 - f_3 v_3}} (\Delta x - v_3 \Delta t) \\ \Delta t'' = \dfrac{1}{\sqrt{1 - f_3 v_3}} (\Delta t - v_3 \Delta x) \end{cases} \quad (19)$$

Using Eqs.(18) and (19) and the arbitrary properties of $\Delta x$ and $\Delta t$, We can further obtain a formula of velocity addition

$$v_3 = \frac{v_1 + v_2}{1 + f_1 v_2} \quad (20)$$

and an important relation

$$f_2 v_1 = f_1 v_2 \quad (21)$$

Eq.(21) can reexpressed as

$$\frac{f_1}{v_1} - \frac{f_2}{v_2} = 0, \text{ i.e., } \frac{f_1}{v_1} = \frac{f_2}{v_2} = K \quad (22)$$

On the other hand, a lot of general physical processes should satisfy quantitative causal relation with no-loss-no-gain character [16,17], e.g., Ref.[18] uses the no-loss-no-gain homeomorphic map transformation satisfying the quantitative causal relation to gain exact strain tensor formulas in Weitzenböck manifold. In fact, some changes ( cause ) of some quantities in (22) must result in the relative some changes ( result ) of the other quantities in (22) so that (22)'s right side keep no-loss-no-gain, i.e., zero, namely, (22) also satisfies the quantitative causal relation and is consistent.

Eq.(22) means that K is a general constant and has nothing to do with any inertial reference system. From(22), we have

$$f_i = K v_i \quad (i = 1, 2) \quad (23)$$

Thus Eqs.(6), (11), (13), (15) and (20) can be rewritten as

$$\frac{\Delta x_1}{\Delta x'} = \frac{\Delta x'}{\Delta x} = \sqrt{1-Kv^2} \tag{24}$$

$$\frac{\Delta t_1}{\Delta t'} = \frac{\Delta t'}{\Delta t} = \sqrt{1-Kv^2} \tag{25}$$

$$\Delta x' = \frac{\Delta x - v\Delta t}{\sqrt{1-Kv^2}} \tag{26}$$

$$\Delta t' = \frac{\Delta t - Kv\Delta x}{\sqrt{1-Kv^2}} \tag{27}$$

$$v_3 = \frac{v_1 + v_2}{1 + Kv_1 v_2} \tag{28}$$

From Eqs.(2) and (22), we know that K's dimension is the inverse of square of velocity dimension

$$[K] = \frac{1}{\left[\frac{L}{T}\right]^2} \tag{29}$$

Using Eqs.(26) and (27) and substituting K with $1/c^2$, the other formula of speed addition can be obtained, which is identical with that in Ref.[12], thus we don't repeat here．

5．**Discussions**

We start only from principle of special relativity, have proved that there exists a general constant K，like the other physical constants，K's value must be determined by means of physical experiments, its possible values are three kinds of likely cases of K ＞ 0, = 0 and ＜ 0. We now discuss them, respectively, as follows:

(i) For K = 0，f = Kv = 0． From (2) we know that the property of the same time has nothing to do with reference system, and Lorentz transformation formulas （26） and （27）are simplified as Galilean transformations, i.e., in the case of  f = 0  without relativistic effects.

(ii) For K > 0, define

$$K = \frac{1}{q^2} \tag{30}$$

From Eqs.（2），（22） and（29）, we know what q has dimension of velocity．
Inserting Eq.（30） into Eqs.（24—27） we have

$$\frac{\Delta x'}{\Delta x} = \sqrt{1 - \frac{v^2}{q^2}} \tag{31}$$

$$\frac{\Delta t'}{\Delta t} = \sqrt{1 - \frac{v^2}{q^2}}$$

$$\Delta x' = \frac{\Delta x - v\Delta t}{\sqrt{1 - \frac{v^2}{q^2}}} \quad (32)$$

$$\Delta t' = \frac{\Delta t - \frac{v}{q^2}\Delta x}{\sqrt{1 - \frac{v^2}{q^2}}} \quad (34)$$

where $v < q$ or $v/q < 1$.

Thus the formula of velocity addition can be rewritten as

$$v_3 = \frac{\frac{v_1}{q} + \frac{v_2}{q}}{1 + \frac{v_1}{q}\frac{v_2}{q}} \quad (35)$$

For
$$v_i < q \quad (i=1,2) \quad (36)$$

we have

$$\frac{v_1}{q} + \frac{v_2}{q} - (1 + \frac{v_1}{q}\frac{v_2}{q}) = (1 - \frac{v_2}{q})(\frac{v_1}{q} - 1) < 0 \quad (37)$$

From Eqs.(35-37), it can be deduced that $v_3 < q$, i.e., the sum $v_3$, in Eq.(35), of the two velocities $v_1$, $v_2 < q$ cannot exceed the speed limit q. If $v_i = q$ (i =1 or 2) or $v_1 = v_2 = q$, then $v_3 = q$, these mean that q is an invariant velocity. Therefore if a kind of matter can move at velocity q, then from the above discussions we have that an observer in any inertial system discovers what the velocity of the matter forever is q, which is just analogous Einstein's invariant principle of light velocity. In this Letter the different aspect is that it is not the premise of building special theory of relativity, but a deduction of principle of special relativity.

And when q in Eqs.(31-35) is substituted by transcending light velocity C and light velocity c, then Eqs. (31-35) and the other discussions about the generalizations to four dimensional space time and so on are identical with those in Ref.[19], which just shows that the classical special theory of relativity in our universe is naturally contained in the general theory of special relativity in cosmology (or multiverse). Thus this Letter proves that the general invariant velocity is originated from the principle of special relativity, namely, discovers the origin of the general invariant velocity; when the general invariant velocity is taken as the general transcending light velocity and the invariant light velocity in current theories, respectively, we get the corresponding special theories of relativity.

(iii) For K < 0, this is another likely case satisfying principle of special relativity, in which the relativistic effects observed would be contrary to the case of K > 0. For instance, from Eqs. (24) and (25) we have that the moving body would be lengthened, moving clock would be quickened, even the value and the direction of $v_3$ in Eq. (28) would be able to be, respectively, infinite and opposite. Now physical experiments point out that our universe chooses the case of K > 0. Whether somewhere does in cosmology (or multiverse) there exist the case of K < 0? We can not give a definite answer up to now, which is a theoretical prediction that possibly exists in the other some places in cosmology (or multiverse). But using the generality of K, even there exists this kind of the universe, it can not connect with our universe by means of coordinate transformation between two inertial systems, because the point of K = 0 is a bifurcation point, through which it gives three types of possible universes in cosmology (or multiverse).

Therefore, we see that light velocity in current theories has very strange properties, e.g., irrespective for the relative observing system is static or motion, the corresponding light velocity is always invariant, namely, it does not satisfy the fundamental plus rule of velocity vectors; and this Letter proves that the strange invariant light velocity in current theories is originated from the principle of special relativity, namely, discovers origin of invariant light velocity.

6. **Summary and Conclusion**

This Letter proves that the general invariant velocity is originated from the principle of special relativity, namely, discovers the origin of the general invariant velocity; when the general invariant velocity is taken as the general transcending light velocity and the invariant light velocity in current theories, respectively, we get the corresponding special theories of relativity. Therefore, we prove that the strange invariant light velocity in current theories is originated from the principle of special relativity, that is, discovers origin of invariant light velocity. Consequently, we, in fact, can see that from now on, the theory of special relativity is just set up on the principle of special relativity, without the strange basic presumption of constant light velocity, which is very essential and important for further development of physics, and the results of this Letter will be utilized and cited broadly because the theory of special relativity is one of the very importantly fundamental theories of modern physics. Furthermore, we demonstrate that there exists a general constant K, for K > 0, = 0 and < 0, they give out three types of possible theories of relative properties in the cosmology.

For K > 0, there exists an invariant velocity $q = \frac{1}{\sqrt{K}}$, Einstein's special relativity is naturally contained in the general theory corresponding to the case of K > 0, and the general theory of special theories of relativity in cosmology just can describe our now world but differing Einstein's special theory of relativity follows:

（i） Here the existence of invariant velocity is not presupposition of the theory but a result of principle of special relativity.

(ii) For K > 0, we deduce that there exists an invariant velocity, but now theory doesn't need that there must be a kind of matter moving at invariant velocity，i.e.，even there not were electromagnetic field that provides a constant light velocity in our universe, the general theory of special theory of relativity in cosmology (or multiverse) still stands，which is different from the special theory of relativity based on the invariant principle of light velocity.

(iii）Although the value of the general constant K still needs to be determined by experiments in this Letter，this is different from testing invariant principle of light velocity by experiments．It is well known that up to now physical experiments have only proved that light velocity is just invariant in a come-and-go path, and people theoretically presume that light velocities in different inertial systems are identical at a higher accuracy, but, with scientific development, this is not able to insure that future experiments would not show that they had minute difference．Deferent from this, it is easy to test  K > 0 because so long as experiments can arrive at a definite accuracy people can give an assertion of this problem, furthermore people can assert the existence of an invariant velocity and may insure that future experiments will not negative the general theory, as to the accurate values of K and q，they are not important for the stand of the general theory.

In order to determine whether K > 0 or < 0, for example, we may measure if life times of moving unstable elementary particles increase with increments of their motion velocity. Now experimental accuracy has given certain results corresponding to K> 0.

The above discussions point out that even on future one day minute difference of light velocities in different inertial systems would be found，which only means that photon has tiny mass and its motion doesn't arrive at the prediction limit velocity, these can not affect whether the general theory of special theory of relativity stands．

For K < 0, this is another likely case satisfying principle of special relativity, in which the moving body would be lengthened, moving clock would be quickened. Because there are different physical constants, different primordial and boundary conditions and the other possible factors in the forming and evolving process of the different parts in cosmology (or multiverse), and using the above investigations, it is deduced that there is the theoretical prediction: there is K < 0 in some places in cosmology (or multiverse)．And using the generality of K，even there exists this kind of the universe，it can not connect with our universe by means of coordinate transformation between two inertial systems, because the point of K = 0 is a bifurcation point, through which it gives three types of possible universes in the cosmology. Finally we obtain the general theory of special theory of relativity in cosmology (or multiverse). Because this Letter predicts that the moving body would be lengthened and moving clock would be quickened in some universes in the cosmology.

However, this Letter cancels the invariant principle of light velocity of Einstein's special theory of relativity, proves that the invariant principle of light velocity and principle of special relativity in Einstein's special relativity are not independent, and demonstrates that the invariant velocities are originated from principle of special relativity, these have very important and practical meanings in knowing the real universe. And Ref.[20] utilizes invariant principle of light velocity and so on to investigate the principle of relativity and the special relativity triple, using this Letter's results one can improve Ref.[20]'s investigation.

Especially, once a kind of matter with the maximally invariant velocity that may is superluminal or just equal to light velocity is determined by current physical experiments, then the invariant velocity can be taken as a special example of the general invariant velocity achieved in this Letter, and then all results of current theories are consistent by utilizing this Letter's theory.

References
[1] M. Aryal and A. Vilenkin, Phys. Lett. B199, 3 (1987)351
[2] A. Linde, Inflation and Quantum Cosmology, Academic Press, 1990.
[3] A. Linde et al, Phys. Rev. D49, 4(1994)1783.
[4] M. J. Rees, Concluding Perspective, astro-ph/0101268, to appear in New Cosmological Data and the Values of the Fundamental Parameters, ed. A. Lasenby & A. Wilkinson (ASP Publications).
[5] A. Linde, Recent Progress in Inflationary Cosmology, astro-ph/9601004, An invited talk at the 1st RESCEU International Symposium on "The Cosmological Constant and the Evolution of the Universe," Tokyo, November 1995.

[6]  J. Stewart, Advanced General Relativity, Cambridge University Press, Cambridge, New York,1990.
[7]  Y．S．Duan, Zh. Eksp. Theor. Fiz.,17（1954）756; Y．S．Duan et al, Nucl. Phys. B699 (2004) 174．
[8]  M．Carmeli, Classical Fields: General Relativity and Gauge Theory, A Wiley-Interscience Publication, New York, 1982; Y. C. Huang et al, Modern Astrophysics and Cosmology, Science Press, in press.
[9]   L.J. Eisenberg, Am. J. Phys., 35 (1967) 649.
[10]  J. H. Field, Am. J. Phys., 69 (2001 ) 569.
[11]  A. Einstein, The meaning of relativity, （1954）28．
[12]  J. H. Field, Helv. Phys. Acta 70 ( 1997 ) 542.
[13]  N. D. Mermin, Am. J. Phys., 52 (1984 ) 119.
[14]  V. Berzi and V. Gorini, J. Math. Phys., 10 (1969) 1518.
[15]  In J. A. Wheeler and R. P. Feynman, Rev. Mod. Phys., 21 (1949 ) 425.
[16] Y. C. Huang, X. G. Lee and M. X. Shao, Modern Physics Letters, A21(2006) 1107; Y. C. Huang and C. X. Yu, Phys. Rev. D 75, 044011 (2007).
[17] Y. C. Huang and Q. H. Huo, Physics Letters, B662(2008) 290; Y. C. Huang and L. X. Yi, Annals of Physics, 325 (2010) 2140; Y. C. Huang, L. Liao and X. G. Lee, The European Physical Journal, C60 (2009) 481.
[18] Y. C. Huang and B. L. Lin, Physics Letters, A299(2002) 644.
[19] J. S. Synge，Relativity: The special relativity, Amsterdam, North Holland Pub., （1964）.
[20] H. Y. Guo, H. T. Wu and B. Zhou, Physics Letters, B670 (2009) 437.